\documentclass[runningheads]{llncs}

\usepackage[hidelinks,draft,bookmarks=false]{hyperref}
\usepackage[T1]{fontenc}
\usepackage{graphicx}
\usepackage{mathrsfs}
\usepackage{amsmath}
\usepackage{amssymb}
\usepackage{booktabs} 
\usepackage{orcidlink}
\usepackage{algorithm}
\usepackage{algpseudocode}
\usepackage{geometry}
\usepackage{xcolor}
\algnewcommand{\LeftComment}[1]{\State \(\triangleright\) #1}

\begin{document}

\title{A Digital Twin-based Multi-Agent Reinforcement Learning Framework for Vehicle-to-Grid Coordination}

\author{Zhengchang Hua\inst{1,2}\orcidlink{0000-0002-3970-6129} \and
Panagiotis Oikonomou\inst{3}\orcidlink{0000-0002-5564-2591} \and
Karim Djemame\inst{2}\orcidlink{0000-0001-5811-5263} \and
Nikos Tziritas \inst{3}\orcidlink{0000-0002-2091-2037} \and
Georgios Theodoropoulos\inst{4,1}\orcidlink{0000-0002-7448-5886}}

\authorrunning{Z. Hua et al.}
\titlerunning{A DT-Based MARL Framework for V2G Coordination}

\institute{Southern University of Science and Technology (SUSTech), Shenzhen, China \\
\and
University of Leeds, Leeds, United Kingdom \\
\and
University of Thessaly, Lamia, Greece\\ 
\and
Research Institute of Trustworthy Autonomous Systems (RITAS), SUSTech, Shenzhen, China\\
}
\maketitle            

\begin{abstract}
The coordination of large-scale, decentralised systems, such as a fleet of Electric Vehicles (EVs) in a Vehicle-to-Grid (V2G) network, presents a significant challenge for modern control systems. While collaborative Digital Twins have been proposed as a solution to manage such systems without compromising the privacy of individual agents, deriving globally optimal control policies from the high-level information they share remains an open problem. This paper introduces Digital Twin Assisted Multi-Agent Deep Deterministic Policy Gradient (DT-MADDPG) algorithm, a novel hybrid architecture that integrates a multi-agent reinforcement learning framework with a collaborative DT network. Our core contribution is a simulation-assisted learning algorithm where the centralised critic is enhanced by a predictive global model that is collaboratively built from the privacy-preserving data shared by individual DTs. This approach removes the need for collecting sensitive raw data at a centralised entity, a requirement of traditional multi-agent learning algorithms. Experimental results in a simulated V2G environment demonstrate that DT-MADDPG can achieve coordination performance comparable to the standard MADDPG algorithm while offering significant advantages in terms of data privacy and architectural decentralisation. This work presents a practical and robust framework for deploying intelligent, learning-based coordination in complex, real-world cyber-physical systems.

\keywords{Digital Twin \and Multi-Agent Reinforcement Learning \and Vehicle-to-Grid}
\end{abstract}

\section{Introduction}
The transition towards renewable energy sources has posed a significant challenge on the management of energy generation and storage in modern power grids. Vehicle-to-Grid (V2G) technology has emerged as a promising solution to tackle these challenges \cite{tan2016integration}. In V2G systems, the electric Vehicles (EVs) can not only be scheduled to charge at valley time to better balance the generation-usage unbalance of power grids, but also work as a virtual, aggregated distributed energy storage to discharge stored energy back to the power grid when there are high demands. This capability enhances grid stability and promotes more usage of renewable energy in power supplies. 

A simple example of a V2G network is shown in Figure~\ref{fig:v2g-nwk}. The V2G network includes basic power grid components, such as different energy sources, both renewable and non-renewable. The energy sources connect to the grid to supply energy for power consumers. Typical power consumers, for example, include residential homes and factories. The EVs connect to the grid by connecting to different charging spots, such as the charging cord at home or public charging stations. Different from traditional grid structures where EVs are only regarded as power consumers, in V2G network, the EVs also have the capabilities to feedback electricity to the grid. This ability enables EVs to function as mobile energy pools, storing excess energy when grid supply exceeds consumer demand and discharging power back to the grid when supply is insufficient \cite{ma2012modeling}. While grids have technologies to dynamically adjust the generation and consumption to avoid imbalance, this process typically involves using fossil fuel power generators, which is non-renewable and less ideal in terms of environmental impacts. V2G offers a very promising solution to boost the usage of renewable energy in traditional power grids.

However, there are several problems that hinder the realisation of V2G systems in real-world power grids. The most significant challenge is the management of such a system with high uncertainty originating from different user behaviours and mobility patterns \cite{liu2013opportunities}. Moreover, there are problems with the scale of the system, as a potential V2G system may easily contain thousands of EVs as autonomous agents. These complexities make V2G coordination a challenging problem for traditional, centralised control methods, necessitating more intelligent and adaptive approaches.

Recent work has increasingly focused on the use of Digital Twins to model and manage large complex systems \cite{10.1007/978-3-031-69583-4_18,9361651,10.1007/978-3-031-52670-1_28}. A Digital Twin is a "live" model of a physical entity, continuously updated with real-time data \cite{grieves2016digital}. This continuous synchronization enables sophisticated "what-if" analysis to identify optimal system actions and assess the potential consequences of human interventions \cite{10.1145/3573900.3591121,10.1145/3635306,9283357}. However, the traditional DT-based approach of creating a single, monolithic Digital Twin for the entire V2G network presents significant disadvantages. For such systems with immense scale, a centralised model will suffer from the inevitable problems of single-point-of-failure, high synchronisation overhead, and may raise serious privacy concerns regarding the data of individual owners. 

Therefore, to fill the gap, the approach of federated collaborative Digital Twin has been proposed \cite{10305745}. While the concept of collaboration Digital Twins provides a robust architectural foundation, it also introduces new challenges. The major challenge is how to effectively coordinate these independent, autonomous entities to achieve system-wide objectives. Even though these frameworks allow DTs to collaborate and build global models, a critical research gap remains: how to derive globally optimal and coordinated actions from the shared information, especially for intelligent Digital Twins powered by Artificial Intelligence, e.g. through Reinforcement Learning (RL). 
RL enables agents to learn complex, decentralised decision-making policies through trial-and-error interaction, allowing them to develop emergent coordination strategies that optimise for long-term, collective goals in dynamic environments.

This article proposes a novel solution to the problem of the collaboration of RL-based intelligent twins. Our proposed approach leverages Centralised Training and Decentralised Execution (CTDE) paradigm to learn globally-aware policies. The core of our contribution is a unique training mechanism that uses a decentralised, collaboratively built global prediction model to assist the centralised training process. This method allows for effective, coordinated learning without the need for the collection of sensitive raw data from individual DTs to a central entity, thereby preserving privacy, reducing communication overhead and achieving a high level of performance.

The contributions of this article is as follows:
\begin{itemize}
\item A novel, hybrid architecture and learning algorithm, Digital Twin Assisted Multi-Agent Deep Deterministic Policy Gradient (DT-MADDPG), that integrates a multi-agent reinforcement learning framework with a collaborative Digital Twin network. The core of this contribution is a simulation-assisted critic mechanism that leverages a collaboratively-built global model to enhance the training of decentralised agents, enabling the learning of complex, coordinated policies.
\item The application and validation of this framework on the complex V2G coordination problem. We demonstrate how our approach effectively manages the high uncertainty and conflicting objectives inherent in V2G systems, providing a scalable and robust solution for balancing grid stability with the needs of individual EV owners.
\end{itemize}

The rest of the paper will be organised as follows: Section~\ref{sec:relatedwork} reviews the related work. Section~\ref{sec:architecture} describes the proposed architecture. Section~\ref{sec:algorithm} details the DT-MADDPG algorithm. Section~\ref{sec:experiments} presents the experimental evaluation. Finally, Section~\ref{sec:conclusion} concludes the paper and outlines future work.

\section{Related Work}
\label{sec:relatedwork}
Solving the complex coordination problem in V2G systems requires drawing upon advances in several distinct research domains. To establish the context for our contribution, this section reviews the relevant literature in three key areas. First, we discuss architectures for Collaborative Digital Twins, which provide the structural foundation for multi-owner systems. Second, we examine algorithms from Multi-Agent Reinforcement Learning, which offer powerful methods for intelligent decision-making. Finally, we survey existing methodologies specifically applied to V2G Coordination.

\subsection{Collaborative Digital Twins}
Although the collaboration among DTs is a relatively new field of research, there are several works that explore the possibility of Digital Twin collaborations. Work by Vergara et al. \cite{10305745,43139e3436ca450aae0690bdcd3df024} proposed Federated Digital Twins (FDT) as an enabling technology for collaborative decision making. Their proposed paradigm allows the interconnections among autonomous DTs in the virtual space, which therefore can leverage the information sharing among the DT network to enhance global decision making. To build such collaborative DTs, a straightforward and intuitive approach is to organise the DTs in a hierarchical manner. Villalonga et al. \cite{villalonga2020local} presented a distributed DT framework using several local DTs to make local decisions and aggregate these decisions at a global DT for scheduling and global decision making. 

To address the limitations of these approaches, particularly regarding data privacy, our previous work introduced the Collaborative Digital Twin Framework \cite{10838936}. This framework enables autonomous DTs to cooperate by sharing high-level information, such as predictions and decisions, rather than sensitive raw data, thus preserving the privacy of individual participants. To enhance the robustness of the system against inaccurate information, the framework also incorporates a trust-based mechanism that evaluates the historical accuracy of each DT's predictions. The effectiveness of this approach was demonstrated in a smart grid-based Virtual Power Plant scenario.

While these frameworks provide the architectural foundation for cooperation, they require an intelligent decision-making engine to achieve complex coordination, a role that can be filled by multi-agent reinforcement learning.

\subsection{Multi-Agent Reinforcement Learning}
The field of Multi-Agent Reinforcement Learning has developed different methodologies to address the challenges of multiple agents, such as the non-stationarity of the environment and the multi-agent credit assignment problem. A dominant paradigm to emerge is Centralised Training with Decentralised Execution, which uses global information during the training phase to capture the global outcomes, while ensuring agents can act independently based on local observations during deployment. Actor-critic methods like MADDPG \cite{lowe2017multi} utilise centralised critics to observe the joint actions of all agents to provide a stable learning signal to decentralised actors, proving effective in mixed cooperative-competitive settings. 
In parallel, value-decomposition methods have focused on cooperative tasks by factorising a team's joint value function. For example, Value-Decomposition Networks \cite{sunehag2017value} directly address credit assignment, decomposing the team value function into agent-wise value functions, therefore solving the problem of spurious rewards. 
A third approach involves learning explicit coordination through communication. Differentiable Inter-Agent Learning \cite{NIPS2016_c7635bfd} creates a differentiable communication channel, allowing agents to learn effective communication protocols end-to-end by passing gradients between them during centralised training.

Building on these paradigms, the Federated Multi-Agent Reinforcement Learning (FMARL) has recently emerged to explicitly address the privacy concerns in centralised training. Chu et al. \cite{9889708} propose a federated deep reinforcement learning framework to manage the charging of plug-in electric vehicles. Their approach coordinates charging tasks across distributed nodes to decentralise control, reduce peak loads and preserve the privacy of owners. Similarly, Qiu et al. \cite{QIU2023120526} integrate federated learning with DDPG method to tackle the privacy challenges in a decentralised energy and carbon trading system for a community of smart buildings. These works demonstrate the growing trend of applying MARL to complex energy systems under privacy constraints, motivating the need for novel architectures that can balance coordination with data privacy.

\subsection{Vehicle-to-Grid Coordination}
There has been many works with different methodologies focusing on V2G coordination in recent years. Zhang et al. \cite{9664795} formulated the V2G coordination problem as a mean field game, and proposed an algorithm to enable the distributed control of vehicles based on the equilibrium strategy. Similarly, Zhou et al. \cite{10500853} presented a game-theoretic model for heterogeneous EVs in V2G services, where the individual decision making process of EVs can be captured by an aggregative game framework, and individual behaviours can be affected by overall population behaviour. 
Beyond game theory, mathematical optimisation is a common approach. Chai et al. \cite{CHAI2023108984} proposed a two-stage optimisation method combining day-ahead and real-time optimisation to tackle the challenges in unplanned user behaviour. By incorporating a real-time adjustment layer into the V2G scheduling, the authors has shown that it is possible to balance the needs between power consumers and EV owners. Yoon et al. \cite{7899469} formulated the charging problem in V2G network as a integer linear programming problem, assuming constant-rate charging with known schedules. Similar approaches can also be seen in the work by Lotfi et al. \cite{LOTFI2024925}, in which the authors proposed an optimal coordinated charging model that formulates the EV coordination as a robust linear programming problem.  

\subsection{Summary}
Our review of the literature reveals a significant research gap at the intersection of these three domains: the need for an adaptive and privacy-preserving coordination mechanism. First, while Collaborative Digital Twin frameworks provide solid infrastructure for privacy-preserving communication, they often rely on simple, pre-defined rules for decision-making and lack a mechanism for learning sophisticated, emergent strategies. Second, while traditional V2G Coordination methods based on optimisation and game theory are powerful, they can be less adaptive to the large complex systems with unpredictable human behaviour. Finally, while state-of-the-art Multi-Agent Reinforcement Learning (MARL) algorithms like MADDPG are highly adaptive, they typically assume direct access to a centralised database of all agents' raw state and action information, which conflicts with the privacy-preserving goals of a distributed DT network.

This paper addresses these gaps by proposing a hybrid architecture that combines the strengths of all three areas. We introduce a MARL algorithm, DT-MADDPG, designed to operate within a collaborative DT framework, leveraging the simulation capabilities of DTs to enable effective, centralised training without direct access to sensitive, raw agent data, thereby providing an adaptive and privacy-preserving solution to the V2G coordination.

\section{Architecture Design}
\label{sec:architecture}

\subsection{Architecture Overview}
\begin{figure}[htbp]
    \centering
    \begin{minipage}[t]{0.43\textwidth}
        \centering
        \includegraphics[width=\linewidth]{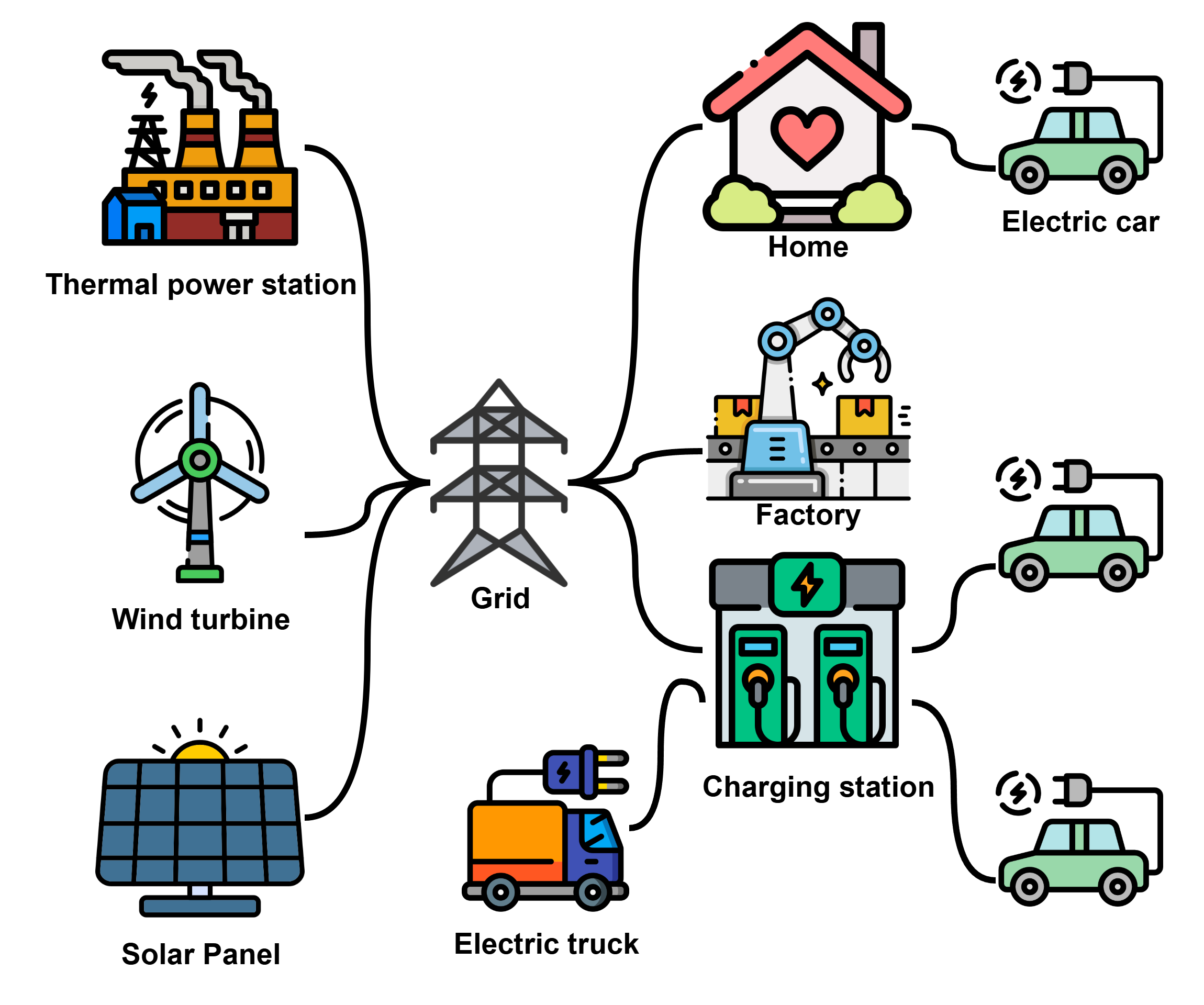}
        \caption{Example V2G Network.}
        \label{fig:v2g-nwk}
    \end{minipage}\hfill
    \begin{minipage}[t]{0.56\textwidth}
        \centering
        \includegraphics[width=\linewidth]{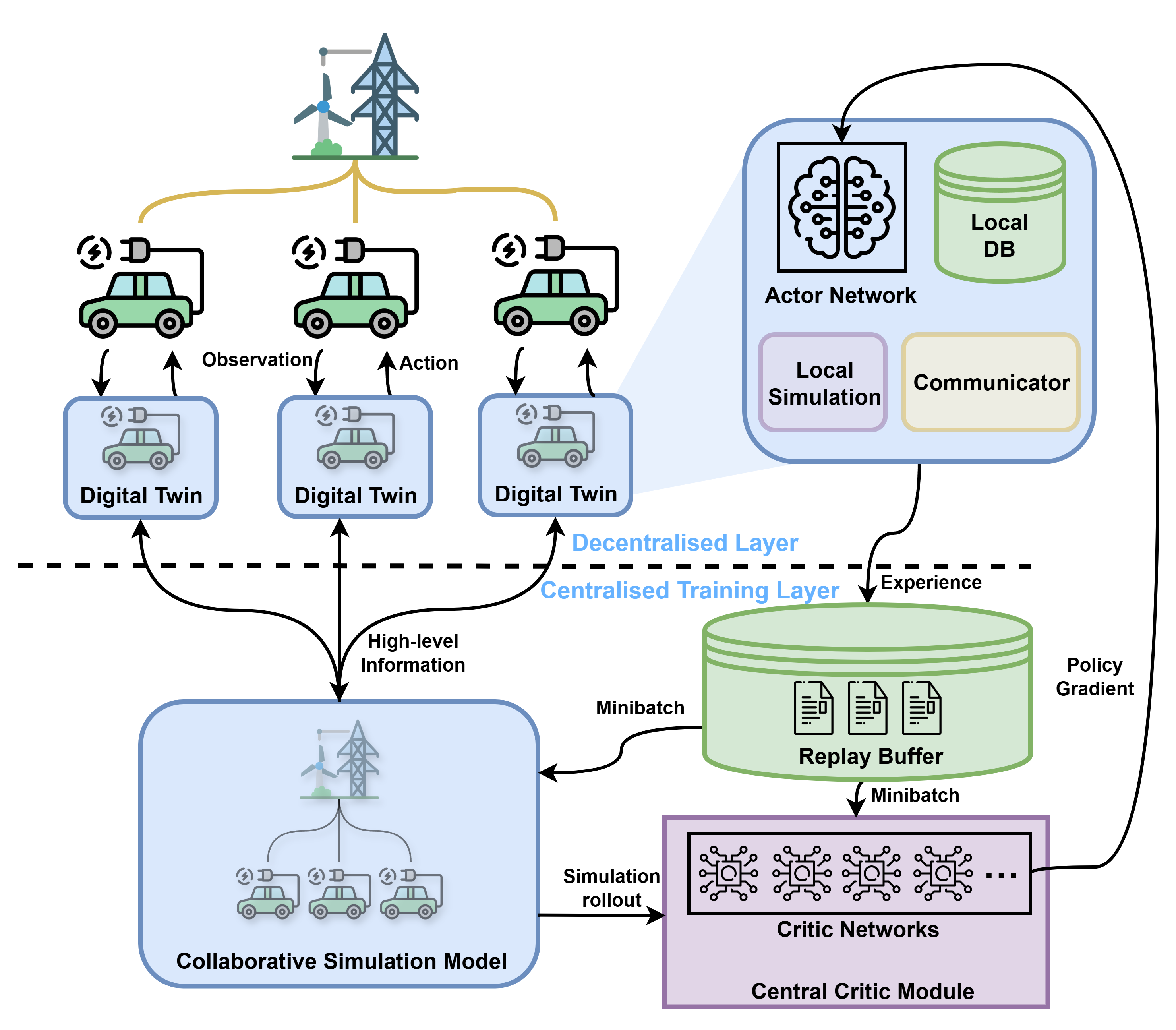}
        \caption{Architecture of the proposed framework.}
        \label{fig:arch}
    \end{minipage}
\end{figure}

Our proposed architecture, as shown in Figure~\ref{fig:arch}, integrates a collaborative DT network with a multi-agent reinforcement learning framework, operating on the paradigm of Centralised Training and Decentralised Execution. In our specific use case, the design is aimed at coordinating a large fleet of EVs in a V2G scenario. The architecture is composed of two primary layers: a decentralised layer of autonomous actors, represented by the EV Digital Twins, and a centralised training layer responsible for training. This structure allows for scalable, real-time decision-making at the agent level while leveraging a global perspective to learn complex and coordinated behaviour.
\vspace{-1em}
\subsubsection{EV Digital Twin}
Each EV in the V2G network has its own local Digital Twin, which acts as an individual, stand-alone node in the Digital Twin Network. This DT contains an integrated decision-maker based on RL, and this decision-maker functions as an agent, or a decentralised actor, in the learning framework.

The core of the Digital Twin's decision maker is a neural network-based actor. This actor takes the EV's local state as input and outputs an action to determine the EV's behaviour, such as its charging/discharging status and the corresponding power rate. Each EV keeps a private instance of its local state, which includes key variables like the current battery level, estimated mileage remaining, and the owner's preferences, all stored in a local database.

Another key component of the local DT is an internal simulator, used for making predictions about its future state. These high-level predictions and decisions, rather than sensitive raw information, are shared within the network via a communicator. This communication enables collaboration while preserving the privacy of each participant.

\vspace{-1em}
\subsubsection{Central Learning Core}
While the actors operate decentrally, they are trained by a centralised learning core that leverages the information shared across the network. This core is responsible for understanding the system-wide impact of joint actions and providing intelligent feedback to the individual actors.

The process begins with the Replay Buffer, an experience buffer that stores experiences collected from all agents interacting with the environment. By sampling minibatches of these experiences, the buffer allows for off-policy learning and breaks the temporal correlations in the data, which is essential for stabilising the training of the deep neural networks.

Data sampled from this buffer then goes into the Collaborative Global Model. This is a high-fidelity simulation model of the V2G system's dynamics, collaboratively constructed from the high-level predictions shared by all the individual EV DTs. Its purpose during training is to receive a state and joint action from the replay buffer and perform a simulation rollout to generate a physics-based value estimate.

This value estimate is then used to assist the Central Critic Module, which learns a global action-value function to assess the quality of a joint policy. Its key innovation is that it learns a residual value on top of the value baseline provided by the Collaborative Global Model's simulation rollout. By learning only the error of the physics-based model, the critic can achieve a more accurate and stable value estimation, which is crucial for guiding the actors toward globally optimal behaviour.

\subsection{System Formulation}
We model the V2G coordination task as a multi-agent collaboration problem, which can be described as a decentralised Partially Observable Markov Decision Process (POMDP). The system consists of $N$ agents, each of them has the goal of learning a policy which maximise its own long-term expected reward.

The formulation of each individual agent $i$ is defined using their state space and action space. Each agent bases its decisions on a local state $s_i$ in its state space $\mathcal{S}$. The local state is a vector composed of the agent's private information and observed environmental data. It includes essential variables such as the current battery level (as ${SoC}_i$), the time remaining before the owner's scheduled departure ($T_{ETD}$), the owner's target SoC at departure (${SoC}_{tgt}$) and the real-time energy price $P_t$. The action space for each agent is a continuous value $a_i$ in the entire action space $\mathcal{A}$ that represents the flow of power in the current time. The power flow is constrained by the maximum charging rate $a_{max,charge}$ and maximum discharging rate $a_{max,discharge}$, therefore $a_i \in [-a_{max,discharge}, a_{max,charge}]$. A negative value corresponds to discharging electricity from the EV into the grid, while a positive value represents charging the EV's battery.

The core components of our learning framework are defined by a set of functions: policy function, action-value function and collaborative global model function. Each agent's behaviour is governed by a deterministic policy $a_i=\pi_{\theta_i}(s_i)$, which is represented by a neural network with parameters $\theta_i$. This function maps agent's local states to a specific charge or discharge action. 
The action-value functions is the centralised critic $Q_i$ with parameters $\phi_i$. It approximated the expected outcome after taking a joint action $a=\{a_1, ..., a_N\}$ in the global state $s$. The key distinction of the proposed architecture is that the critic is enhanced by the prediction generated by the collaborative global model, denoted as $s_{sim}$. Therefore, the critic will be formulated as $Q_i=(s,a,s_{sim}) \rightarrow \mathbb{R}$.
The collaborative global model function $f_{sim}$ represents the global simulation model of the V2G network, which is collaboratively constructed by all participating DTs. It maps the current global state and a joint action to a predicted next state $f_{sim}(s,a) \rightarrow s_{sim}$.

\subsection{Reward Function Design}

To address the complex trade-offs in the V2G network, we designed a hierarchical reward structure that balances system-level goals with the individual preferences of users. The final reward signal for an individual agent $i$ at each time step $t$, denoted as $r_{i}(t)$, is a weighted combination of a shared global reward and the agent's own local reward. This is formally expressed as:

\begin{equation}
    \label{eq:h-reward}
    r_i(t) = w_{global} \times R_{global}(t) + (1-w_{global}) \times R_{local,i}(t)
\end{equation}

In this formulation, $w_{global}$ is a hyper-parameter that controls the agent's alignment with the collective goal. The global reward component, $R_{global}(t)$, is shared among all agents and is derived from system-wide performance metrics. Specifically, it is designed to reward grid stability (minimizing the variance of power drawn from non-renewable sources) and maximizing the use of renewable energy. It is calculated as:

\begin{equation}
    \label{eq:global-reward}
    R_{global}(t) = -\alpha \cdot \text{Var}(P_{grid}(t)) + \beta \cdot \frac{P_{used\_renewable}(t)}{P_{available\_renewable}(t)}
\end{equation}

$\text{Var}(P_{grid}(t))$ is the variance of non-renewable power drawn from the main grid, and the second term is the utilization ratio of available renewable power. The coefficients $\alpha$ and $\beta$ are used to scale these two objectives. The local reward component, $R_{local,i}(t)$, represents the individual objectives of agent $i$.

Individual users typically pursue two distinct and competing goals: satisfaction of their State-of-Charge (SoC) needs and financial revenue. To model this, the local reward is a weighted sum of two terms. We use a single weight, $w_{SoC}$, to model a user's preference, with the constraint that the weights sum to one. The local reward for agent $i$ is shown in Equation~\ref{eq:local-reward}:

\begin{equation}
    \label{eq:local-reward}
    R_{local,i}(t) = w_{SoC} \times R_{SoC}(\text{terminal}) + (1-w_{SoC}) \times R_{revenue}(t)
\end{equation}

The first term, $R_{SoC}$, is a large terminal reward applied only at the agent's scheduled departure time. It is a large positive constant, $C_{success}$, if the final SoC has reached the user's desired threshold, and a large negative penalty, $-C_{fail}$, otherwise. The second term, $R_{revenue}(t)$, is the immediate financial outcome, calculated as:

\begin{equation}
    \label{eq:revenue-reward}
    R_{revenue}(t) = \text{Price}(t) \times P_{flow}(t)
\end{equation}

$\text{Price}(t)$ is the real-time electricity price, and $P_{flow}(t)$ is the power flow of the EV, which is positive for discharging (selling) and negative for charging (buying). This two-level reward structure provides a flexible and powerful mechanism for guiding the agents toward sophisticated, cooperative behaviours that respect both global system needs and individual user priorities.

\begin{table}
\centering
\caption{Table of Notations}
\label{tab:notations}
\begin{tabular}{l@{\hskip 0.5in}l}
\toprule
\textbf{Notation} & \textbf{Description} \\
\midrule
$N$ & Total number of agents (EVs) in the system. \\
$i$ & Index for an individual agent, where $i \in \{1, \dots, N\}$. \\
$s$ & Global state of the environment, composed of all local observations. \\
$o_i$ & Local observation for agent $i$. \\
$a$ & Joint action, composed of all individual agent actions $\{a_1, \dots, a_N\}$. \\
$a_i$ & Action taken by agent $i$. \\
$r_i$ & Reward received by agent $i$. \\
$\mathcal{N}$ & Random noise to encourage exploration.\\
$\mathcal{D}$ & Central replay buffer, which stores experience tuples. \\
$\mu_i(\theta_i)$ & Actor network for agent $i$, parametrised by $\theta_i$. \\
$Q_i(\phi_i)$ & Complete action-value function for agent $i$. \\
$Q_{res,i}(\phi_i)$ & Residual critic network for agent $i$, parametrised by $\phi_i$. \\
$f_{sim}$ & Collaborative Global Model, which functions as a system simulator. \\
$R_{sim}$ & Value baseline calculated from the simulation rollout. \\
$y_i$ & Target value used to train the critic for agent $i$. \\
$\mu'_i$, $Q'_{res,i}$ & Target networks for the actor and residual critic, respectively. \\
$R_{global}(t)$ & Shared global reward component at time $t$. \\
$R_{local,i}(t)$ & Local reward component for agent $i$ at time $t$. \\
$R_{revenue}(t)$ & Immediate economic reward component at time $t$. \\
$R_{SoC}(\text{terminal})$ & Terminal reward for SoC satisfaction at the end of a session. \\
$P_{grid}(t)$ & Power drawn from the non-renewable grid source at time $t$. \\
$P_{flow}(t)$ & Power flow of an EV at time $t$ (positive for discharging, negative for charging). \\
$\text{Price}(t)$ & Real-time price of electricity at time $t$. \\
$C_{success}, C_{fail}$ & Positive constants for the terminal SoC success reward and failure penalty. \\
$w_{global}$ & Weight balancing the global and local components of the reward function. \\
$w_{soc}$ & Weight balancing SoC satisfaction and revenue in the local reward function. \\
$\gamma$ & Discount factor for future rewards. \\
$\tau$ & Soft update parameter for updating the target networks. \\
$k'$ & Simulation rollout horizon (number of steps). \\
\bottomrule
\end{tabular}
\end{table}

\section{DT-MADDPG Algorithm}
\label{sec:algorithm}
To address the cooperative control problem in a multi-agent environment like the V2G system, we propose the \textbf{Digital Twin-Assisted Multi-Agent Deep Deterministic Policy Gradient} (DT-MADDPG) algorithm. Our algorithm builds upon the MADDPG framework but introduces a novel learning mechanism that deeply integrates the predictive power of the collaborative Digital Twin network. The core innovation is a simulation-assisted critic that utilises a value decomposition approach, separating the physics-based, short-term value from the learned, long-term residual value. This structure grounds the learning process, reduces the critic's learning burden, and leads to more stable and efficient policy optimisation.

The fundamental premise of our approach is that simulation models are highly effective at making accurate, high-fidelity predictions over a short-term horizon, but their accuracy can degrade over longer time frames due to compounding uncertainties and complex emergent behaviours. In contrast, reinforcement learning critics excel at learning long-term strategic values but can be sample-inefficient. Therefore, our architecture is explicitly designed to leverage the strengths of both. We use the simulation to accurately calculate the predictable, short-term value baseline ($R_{sim}$), and assign the neural network critic the task of learning the residual part ($Q_{res}$), which corresponds to the more complex, long-term value beyond the simulation's reliable horizon.

Our proposed algorithm differs from MADDPG by leveraging the capabilities of the collaborative Digital Twin framework. The collaborative framework is able to construct a global simulation model that enables the short-term prediction of the system states. However, relying only on these immediate predictions is insufficient for learning optimal, far-sighted policies, as it fails to capture the complex, long-term strategic consequences of an action. Instead of using a simple one-step prediction, we employ a more sophisticated mechanism involving two key steps: \textbf{Simulation Rollout} and \textbf{Value Decomposition}. For a given state-action pair $(s,a)$ sampled from the replay buffer, we use the global simulation model $f_{sim}$ to perform a short-term simulation rollout for the next $k$ timestamps. This process will generate a predicted trajectory of k pairs of future states and rewards $(s_{t+1}, r_{t+1}, \dots, s_{t+k}, r_{t+k})$. We then use the simulated trajectory to decompose the total Q-value into two parts. Firstly, we calculate the short-term estimation of a specific action by finding the discounted sum of the predicted rewards from the rollout, denoted in Equation~\ref{eq:short-term},where $\gamma$ is the discount factor. This reward models the short-term, model-based estimate of the short-term value.

\vspace{-1em}
\begin{center}
    \begin{minipage}{0.4\textwidth}
        \centering
        \begin{equation}
                \label{eq:short-term}
                R_{sim}(s,a) = \sum_{j=1}^{k}\gamma^{j-1}r_{t+j}
        \end{equation}
    \end{minipage}
    \hfill
    \begin{minipage}{0.54\textwidth}
        \centering
        \begin{equation}
            \label{eq:action-value}
            Q_i(s,a)=R_{sim}(s,a) +  Q_{res,i}(s,a)
        \end{equation}
    \end{minipage}

\end{center}

Since the accuracy of the simulation model tends to decay in long-term predictions, we reframe the critic network's task to learn the residual value function $Q_{res,i}$, which captures the long-term value beyond the horizon of simulation. Therefore, the complete action-value function for the agent $i$ is the sum of these two components, as in Equation~\ref{eq:action-value}.

Based on this value decomposition, the critic for each agent $i$ is updated by minimizing the loss between the target value $y_i$ and the composite Q-value.
\begin{equation}
    \label{eq:loss}
    L(\phi_i)=\mathbb{E}[(y_i-(R_{sim}(s,a)+Q_{res,i}(s,a)))^2]
\end{equation}
The target value $y_i$ is calculated using the target networks, as is standard in actor-critic methods. The actor policy for each agent is then updated using the deterministic policy gradient derived from the full composite Q-function, guiding the agent to take actions that maximise both the simulated short-term return and the learned long-term residual value.

The complete training procedure for our proposed DT-MADDPG algorithm is summarised in Algorithm 1. The process begins by initializing the actor network $\mu_i$ and the residual critic network $Q_{res,i}$ with random weights for each of the $N$ agents. Their corresponding target networks, $\mu'_i$ and $Q'_{res,i}$, are then created by copying the weights from the main networks. A replay buffer $\mathcal{D}$ is initialised to store experiences, and the Collaborative Global Model, $f_{sim}$, which represents the simulation capabilities of the DT network, is also initialised.

The algorithm proceeds in episodes, with each episode consisting of multiple time steps where agents interact with the environment to collect data. In each step, every agent $i$ uses its local observation $o_i$ and its actor policy $\mu_i$ to select a deterministic action $a_i$. To encourage exploration, random noise $\mathcal{N}_t$ is added to this action. The joint action $a = \{a_1, \dots, a_N\}$ from all agents is then executed in the environment, causing a transition to a new global state $s'$ and yielding a reward $r$. The complete experience tuple $(s, a, r, s')$ is stored in the central replay buffer $\mathcal{D}$.

\begin{algorithm}[hp]
\caption{Digital Twin-Assisted MADDPG (DT-MADDPG)}
\label{alg:dt-maddpg}
\begin{algorithmic}[1]
\State \textbf{Input:} Number of agents $N$, simulation rollout horizon $k'$, hyper-parameters $\gamma, \tau$.
\State \textbf{Output:} Trained decentralised actor policies $\{\mu_1, \dots, \mu_N\}$.
\Statex
\State \textbf{Initialise} for each agent $i=1, \dots, N$:
\State \quad Actor network $\mu_i$ with parameters $\theta_i$.
\State \quad Residual critic network $Q_{res,i}$ with parameters $\phi_i$.
\State \quad Target networks $\mu'_i$ and $Q'_{res,i}$ with weights $\theta'_i \leftarrow \theta_i$ and $\phi'_i \leftarrow \phi_i$.
\State \textbf{Initialise} replay buffer $\mathcal{D}$ and Collaborative Global Model $f_{sim}$.

\For{episode = 1 to M}
    \State Receive initial observations $\{o_1, \dots, o_N\}$.
    \For{t = 1 to max\_steps}
        \State For each agent $i$, select action $a_i = \mu_i(o_i) + \mathcal{N}_t$.
        \State Let joint action be $a = \{a_1, \dots, a_N\}$.
        \State Execute $a$, observe rewards $r = \{r_1, \dots, r_N\}$ and next observations $\{o'_1, \dots, o'_N\}$.
        \State Store the tuple $(\{o_i\}, \{a_i\}, \{r_i\}, \{o'_i\})_{i=1..N}$ in the replay buffer $\mathcal{D}$.
        
        \If{replay buffer $\mathcal{D}$ contains enough samples}
            \State Sample a random minibatch of $K$ experiences.
            \State Let a sample $k$ be $(\{o_i^k\}, \{a_i^k\}, \{r_i^k\}, \{o'_{i^k}\})$.
            
            \LeftComment{Perform simulation rollout using joint information}
            \State Calculate value baselines $R_{sim}^k = \text{Rollout}(f_{sim}, \{o_i^k\}, \{a_i^k\}, \text{horizon}=k')$.
            
            \For{agent $i = 1$ to $N$}
                \LeftComment{Update critic for agent i using global information}
                \State Form the target joint action: $a'^k = \{\mu'_1(o'^k_1), \dots, \mu'_N(o'^k_N)\}$.
                \State Calculate target simulation baseline: $R_{sim, next}^k = \text{Rollout}(f_{sim}, \{o'^k_i\}, a'^k, \text{horizon}=k')$.
                \State Set target value: $y_i^k = r_i^k + \gamma (R_{sim, next}^k + Q'_{res, i}(\{o'^k_j\}_{j=1..N}, a'^k))$.
                \State Update critic by minimizing the loss: 
                \Statex \quad\quad\quad $\mathcal{L}(\phi_i) = \frac{1}{K}\sum_k \left(y_i^k - (R_{sim}^k + Q_{res,i}(\{o_j^k\}_{j=1..N}, \{a_j^k\}_{j=1..N}))\right)^2$.
                
                \State Update actor for agent $i$ using the sampled policy gradient w.r.t $Q_i$.
            \EndFor
            
            \LeftComment{Soft update all target networks}
            \For{agent $i = 1$ to $N$}
                \State $\phi'_i \leftarrow \tau \phi_i + (1-\tau)\phi'_i$
                \State $\theta'_i \leftarrow \tau \theta_i + (1-\tau)\theta'_i$
            \EndFor
        \EndIf
    \EndFor
\EndFor
\end{algorithmic}
\end{algorithm}

The centralised training phase begins once the replay buffer contains a sufficient number of experiences. A random minibatch of $K$ experiences is sampled from the buffer to update all networks. For each experience $(s_k, a_k)$ in the minibatch, the collaborative global model $f_{sim}$ performs a short-horizon simulation rollout. This process calculates $R^k_{sim}$, a model-based estimate of the short-term value, by computing the discounted sum of rewards predicted by the simulation. The algorithm then iterates through each agent to update its networks. The residual critic, $Q_{res,i}$, is trained to capture the long-term value that the simulation cannot predict. This is achieved by minimizing the loss between a stable target value, $y_i^k$ (computed using the target networks), and the critic's predicted composite value, which is the sum $(R^k_{sim} + Q_{res,i}(s^k,a^k))$. Subsequently, each actor $\mu_i$ is updated using the policy gradient derived from its corresponding composite Q-function, $Q_i = R_{sim} + Q_{res,i}$. Finally, the weights of all target networks are updated via a soft update controlled by the parameter $\tau$ to stabilise the learning process.

\section{Experimental Evaluation}
\label{sec:experiments}

\subsection{Experiment Setup}
To validate the effectiveness of our proposed DT-MADDPG architecture, we designed a series of experiments within a simulated V2G environment. The objective is to assess the ability of our framework to learn effective coordination policies by comparing its performance on several key metrics against established baseline methods.

We developed a discrete-event based simulation representing a local power distribution grid servicing a community with a fleet of EVs. The simulator is programmed in Python using the SimPy library and the objects in the simulation such as EV, generators and consumers are modelled as SimPy processes. The environment consists of a power grid model and the EV fleet model. The grid has a fluctuating base load from residential and commercial uses and two primary power sources: a main grid connection representing non-renewable (fossil fuel) power, and a local renewable energy source (e.g., solar or wind farms) with variable output based on historical data. In these settings, we assume that the load of the grid can be balanced by adjusting the output power generated by non-renewable resources. A dynamic pricing mechanism dictates the cost of drawing power from and the revenue from selling power back to the main grid. The dynamic pricing is segmented, with three periods - Peak, Valley and Normal. The detailed price is listed in Table~\ref{tbl:elec-price}. The data is taken from the residential electricity prices in the city of Shenzhen, China. We also assume the price of selling electricity back to the grid is the same as the energy price in the given period.

\begin{table}[htbp]
\centering
\label{tbl:elec-price}
\caption{Dynamic Electricity Price}
\begin{tabular}{|c|c|c|}
\hline
Period & Time               & Price (CNY/kWh) \\ \hline
Peak   & 10-12; 14-19       & 1.1121          \\ \hline
Normal & 8-10; 12-14; 19-24 & 0.6542          \\ \hline
Valley & 0-8                & 0.2486          \\ \hline
\end{tabular}
\end{table}

In the simulation, a fleet of $N$ EVs, each represented by a Digital Twin agent as described in Section~\ref{sec:architecture} is placed in the grid environment. Each EV is assigned a unique battery capacity, charging/discharging efficiency, and a stochastic behavioural profile governing its daily arrival time, departure time, and energy consumption. To study the scalability of the system. we have also constructed a model of the network infrastructure. While there are no constraints on the bandwidth or latency, it is still worth studying the communication overhead of different approaches. For the network model, we used Barabási–Albert model \cite{barabasi1999emergence}. The experiments are conducted on a bare-metal server with 4 Intel Xeon Gold 6230 CPU and a NVIDIA RTX A6000 GPU.

We evaluate the performance of all approaches using a combination of scenario-specific and system-level metrics:
\begin{itemize}

\item Grid Stability: The primary goal of V2G is to stabilise the grid by reducing reliance on fossil fuels during peak demand (peak shaving) and absorbing excess energy during off-peak times (valley filling). We visualise this by plotting the power drawn from the non-renewable source over time.

\item Renewable Energy Utilisation: This metric measures the percentage of available renewable energy that is successfully used by the system, either consumed directly or stored in EV batteries and consumed in the future, instead of being disposed of. This will be presented as a bar chart, where higher utilisation is desirable.

\item User Satisfaction: We measure the framework's ability to meet the needs of individual EV owners via the Owner Goal Success Rate. This is the percentage of EVs that successfully meet their owner-defined State-of-Charge (SoC) requirements by their scheduled departure time. Results will be compared across algorithms using a bar chart.

\item User Revenue: To assess the economic benefit for participants, we calculate the Aggregated User Revenue. This is the total income earned by all EV owners from selling energy back to the grid, minus their total charging costs. This will also be presented in a comparative bar chart.

\item Communication Overhead: The efficiency of the underlying collaborative framework is measured by the average number of messages exchanged between DTs per simulation step. Also, to showcase the potential network congestion, we plot the distribution of network load (average number of message flowing through a network node) to reveal the network usage.
\end{itemize}

We evaluate the performance of our proposed DT-MADDPG algorithm against two distinct MARL baselines to demonstrate its advantages:
\begin{itemize}
\item Independent Learners (IL): In this approach, each EV agent is a fully decentralised reinforcement learner with its own actor and critic. It learns a policy based solely on its local observations, representing a non-collaborating but still intelligent baseline.

\item Standard MADDPG: This is the canonical centralised training with decentralised execution algorithm. A central critic collects state and action information from all agents to guide training. This baseline is the crucial ablation study, allowing us to isolate and measure the performance gain attributable directly to our novel simulation-assisted critic.
\end{itemize}

\subsection{Results and analysis}

\begin{figure}[htbp]
    \centering
    \begin{minipage}{0.49\textwidth}
        \centering
        \includegraphics[width=\linewidth]{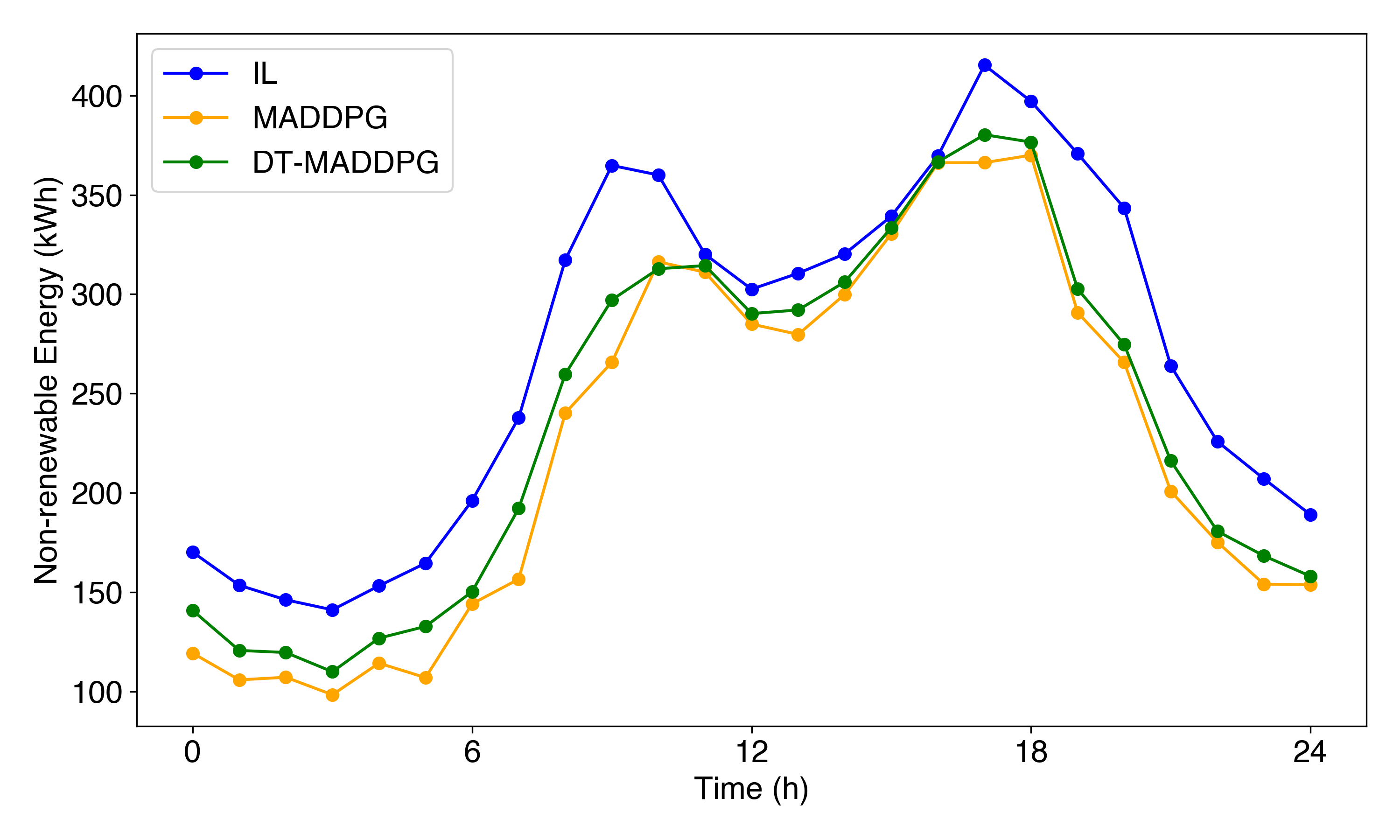}
        \caption{Energy drained from non-renewable energy sources, indicating grid stability.}
        \label{fig:stability}
    \end{minipage}\hfill
    \begin{minipage}{0.49\textwidth}
        \centering
        \includegraphics[width=\linewidth]{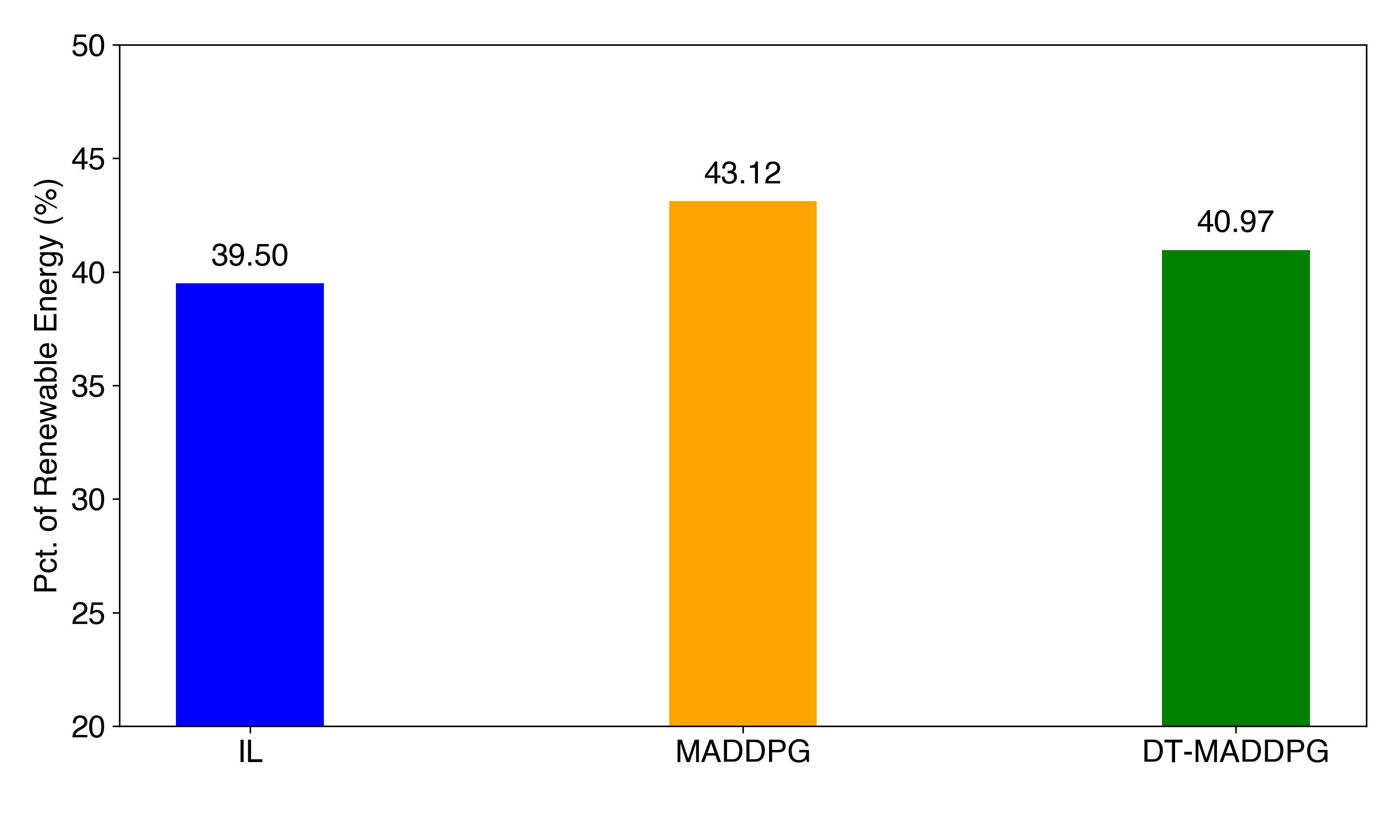}
        \caption{Percentage of renewable energy utilisation with different algorithms.}
        \label{fig:utilization}
    \end{minipage}

\end{figure}

First, we assessed the core performance of the models in managing the power grid. Figure~\ref{fig:stability} illustrates grid stability by plotting the required fossil fuel generation (Y-axis) over a 24-hour period (X-axis). The results clearly show that both MADDPG and DT-MADDPG outperform the IL baseline by requiring less fossil fuel. Crucially, DT-MADDPG achieves stability comparable to the fully centralised MADDPG, proving its effectiveness in coordinating the V2G network.

This high performance extends to the use of renewables, as shown in Figure~\ref{fig:utilization}. DT-MADDPG achieves a renewable energy utilisation of 40.97\%, a result that is nearly identical to MADDPG and significantly higher than IL. These findings establish that DT-MADDPG maintains top-tier performance on primary system objectives without requiring full centralisation.

\begin{figure}[htbp]
    \centering
    \begin{minipage}{0.49\textwidth}
        \centering
        \includegraphics[width=\linewidth]{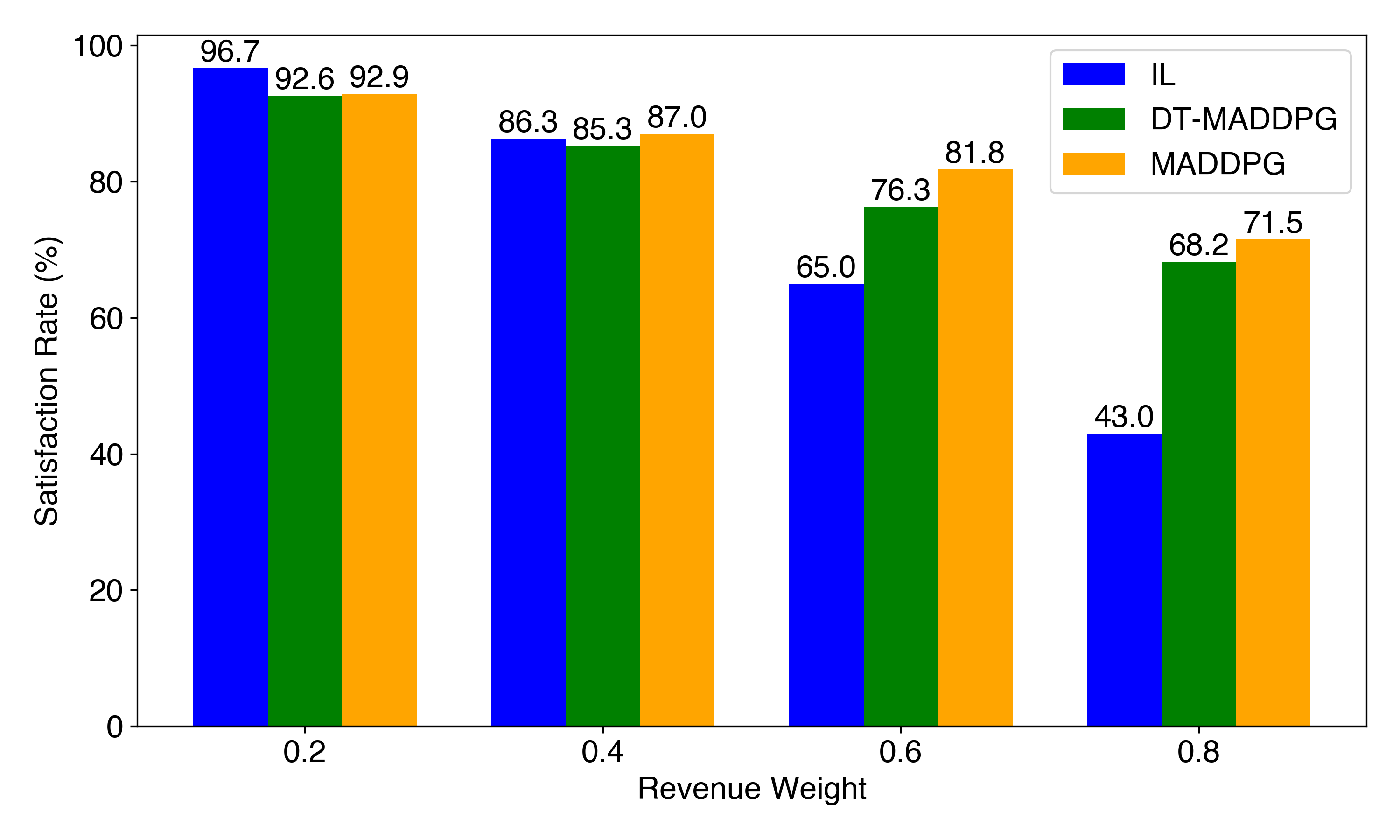}
        \caption{User satisfaction rate as a function of the revenue preference weight ($w_{revenue}$).}
        \label{fig:satisfaction}
    \end{minipage}\hfill
    \begin{minipage}{0.49\textwidth}
        \centering
        \includegraphics[width=\linewidth]{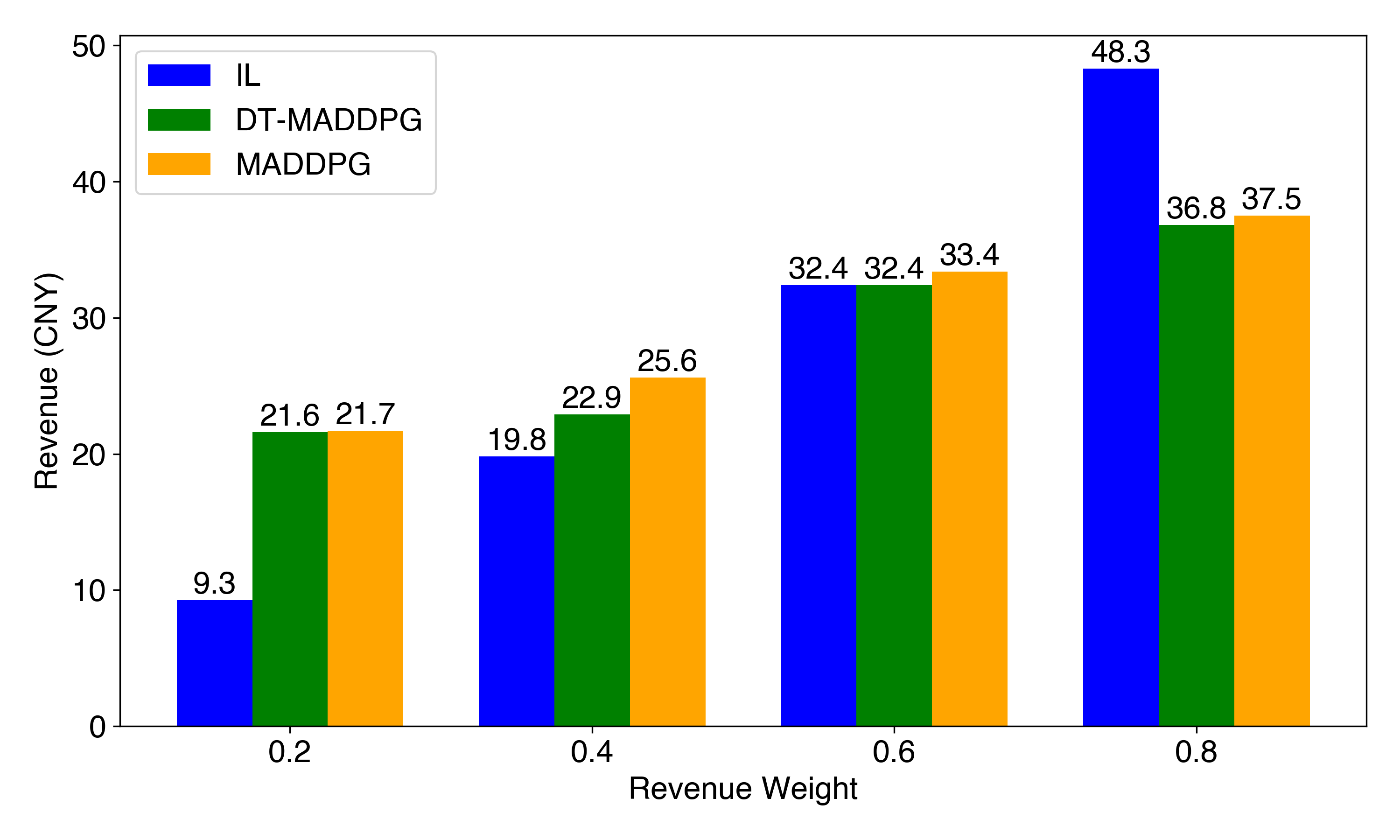}
        \caption{Aggregated user revenue as a function of the revenue preference weight ($w_{revenue}$)}
        \label{fig:revenue}
    \end{minipage}

\end{figure}

To investigate how the models adapt to different user priorities, we conducted a sensitivity analysis by varying the revenue weight, $w_{revnue}$, from 0.2 to 0.8. A higher $w_{revnue}$ encourages agents to prioritise revenue generation over maintaining their target State of Charge.
As expected, increasing this weight leads to a clear trade-off: user satisfaction declines (Figure~\ref{fig:satisfaction}) while user revenue increases (Figure~\ref{fig:revenue}) for all methods. However, a key trend emerges: the IL baseline is highly sensitive to this parameter change, showing volatile swings in performance. In contrast, MADDPG and DT-MADDPG demonstrate significantly more stability. This resilience can be attributed to their coordination mechanisms; while IL agents narrowly focus on local, weight-sensitive goals, MADDPG and DT-MADDPG agents must balance these with the global objective of grid stability. This moderates their response to changes in local incentives, leading to more robust and predictable system behaviour.

\begin{figure}[htbp]
    \centering
    \begin{minipage}{0.49\textwidth}
        \centering
        \includegraphics[width=\linewidth]{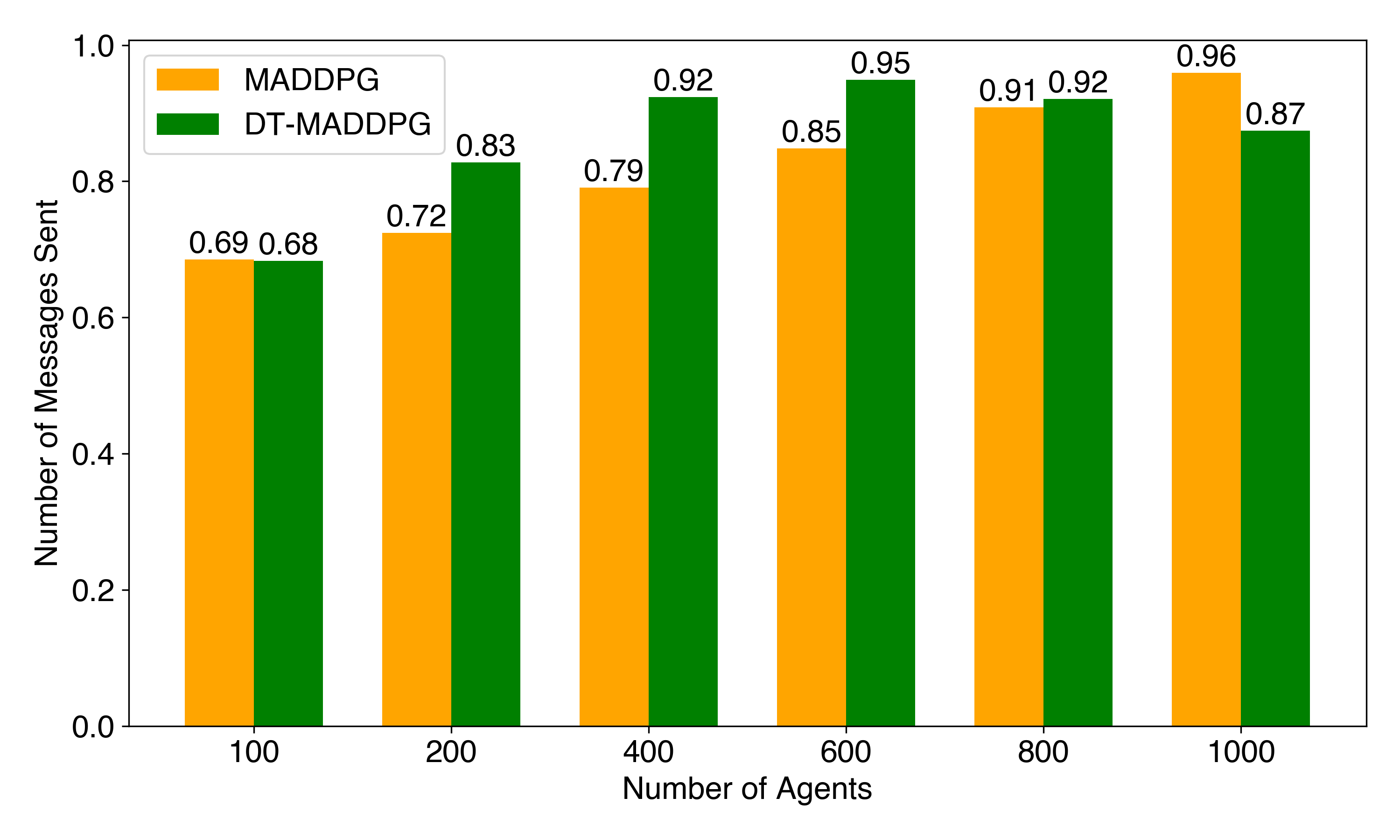}
        \caption{Scalability of communication overhead versus the number of agents.}
        \label{fig:num_messages}
    \end{minipage}\hfill
    \begin{minipage}{0.49\textwidth}
        \centering
        \includegraphics[width=\linewidth]{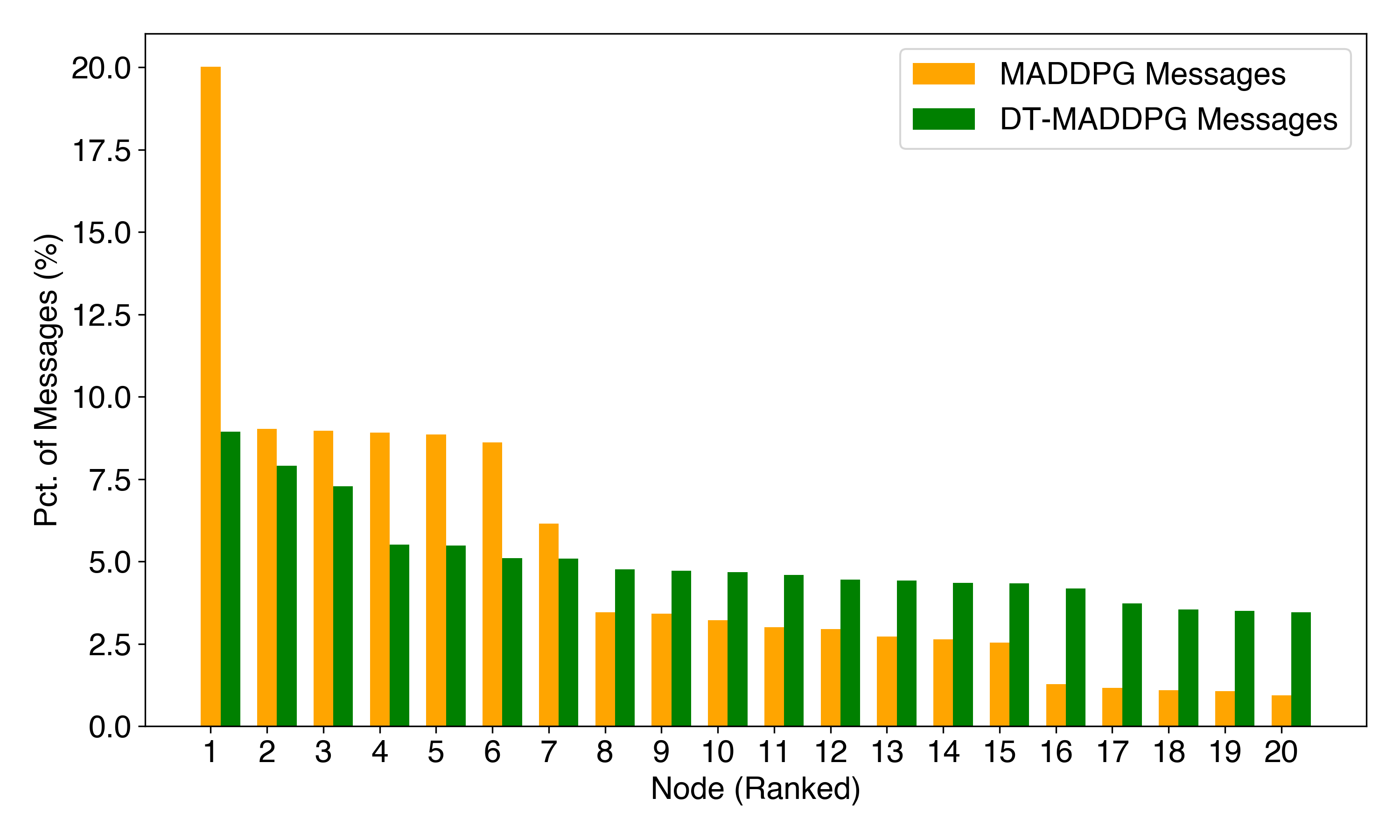}
        \caption{Distribution of network communication load across nodes.}
        \label{fig:message_distribution}
    \end{minipage}

\end{figure}

While MADDPG's access to complete global data may theoretically yield slightly better performance, the primary advantage of DT-MADDPG lies in its privacy-preserving architecture and communication efficiency. At first glance, the total communication volume appears similar. Figure~\ref{fig:num_messages} shows that the average number of messages per agent scales nearly linearly and at a comparable rate for both MADDPG and DT-MADDPG as the system grows. However, the critical advantage of DT-MADDPG is revealed in the distribution of this network traffic, shown in Figure~\ref{fig:message_distribution}. This figure plots the percentage of total number of messages passing through each network node. MADDPG's centralised architecture creates significant network imbalances, resulting in bottlenecks where a few nodes handle a very high load. In contrast, DT-MADDPG's decentralised approach where most data is exchanged locally between DTs leads to a substantially flatter and more balanced network load. This demonstrates that DT-MADDPG achieves its strong coordination performance without the communication bottlenecks of centralised systems, making it a far more robust and scalable solution for real-world deployment.

\section{Conclusion and Future Work}
\label{sec:conclusion}
In this paper, we introduced DT-MADDPG, a novel hybrid architecture that addresses the challenge of coordinating large-scale V2G systems. Our core contribution is demonstrating that a decentralised, privacy-preserving framework can achieve strong coordination performance without the communication bottlenecks inherent in centralised systems. While achieving performance comparable to the standard MADDPG, our results show that DT-MADDPG's primary advantage lies in its ability to create a balanced and distributed network load. This architectural feature makes it a significantly more robust and scalable solution for real-world smart grid applications.

We acknowledge several limitations that provide directions for future research. First, this study relies on simulation-only validation and assumes accurate simulation models, which does not account for real-world factors like hardware variability or sensor noise. Second, our model assumes homogeneous user profiles and preferences, whereas real-world heterogeneity could introduce new coordination dynamics. Finally, while our framework is designed to be privacy-preserving, this study does not provide a formal, quantitative analysis of privacy leakage, which is necessary for a thorough discussion of the privacy-utility trade-off.

Building on these limitations, future work will focus on extending this framework by formally incorporating privacy as a quantifiable penalty in the reward function. By adding a negative reward term proportional to privacy loss, agents will be forced to learn an explicit trade-off, optimizing their policies to find a Pareto-optimal balance between maximizing performance and minimizing data disclosure. Further avenues include developing more complex economic models, such as dynamic energy pricing based on supply and demand or auction-based systems for resource allocation. These models would move beyond simple transactions to create a true digital marketplace with emergent economic behaviours. Finally, we plan to validate the learned policies on physical hardware testbeds to bridge the sim-to-real gap.

\section*{Acknowledgements}
This work was supported in part by the Research Institute of Trustworthy Autonomous Systems (RITAS), Shenzhen Science and Technology Program (No. GJHZ20210705141807022), Guangdong Province Innovative and Entrepreneurial Team Programme (No. 2017ZT07X386), and SUSTech-University of Leeds Collaborative PhD Programme. Georgios Theodoropoulos is the corresponding author of this article.

\bibliographystyle{splncs04}
\bibliography{ref}

\begin{thebibliography}{10}
\providecommand{\url}[1]{\texttt{#1}}
\providecommand{\urlprefix}{URL }
\providecommand{\doi}[1]{https://doi.org/#1}

\bibitem{barabasi1999emergence}
Barab{\'a}si, A.L., Albert, R.: Emergence of scaling in random networks. science  \textbf{286}(5439),  509--512 (1999)

\bibitem{CHAI2023108984}
Chai, Y.T., Che, H.S., Tan, C., Tan, W.N., Yip, S.C., Gan, M.T.: A two-stage optimization method for vehicle to grid coordination considering building and electric vehicle user expectations. International Journal of Electrical Power \& Energy Systems  \textbf{148},  108984 (2023). \doi{10.1016/j.ijepes.2023.108984}

\bibitem{9889708}
Chu, Y., Wei, Z., Fang, X., Chen, S., Zhou, Y.: A multiagent federated reinforcement learning approach for plug-in electric vehicle fleet charging coordination in a residential community. IEEE Access  \textbf{10},  98535--98548 (2022). \doi{10.1109/ACCESS.2022.3206020}

\bibitem{10.1145/3573900.3591121}
Diamantopoulos, G., Bahsoon, R., Tziritas, N., Theodoropoulos, G.: Symbchainsim: A novel simulation tool for dynamic and adaptive blockchain management and its trilemma tradeoff. In: Proceedings of the 2023 ACM SIGSIM Conference on Principles of Advanced Discrete Simulation. p. 118–127. SIGSIM-PADS '23, Association for Computing Machinery, New York, NY, USA (2023). \doi{10.1145/3573900.3591121}

\bibitem{10.1007/978-3-031-52670-1_28}
Diamantopoulos, G., Tziritas, N., Bahsoon, R., Theodoropoulos, G.: Dynamic data-driven digital twins for blockchain systems. In: Blasch, E., Darema, F., Aved, A. (eds.) Dynamic Data Driven Applications Systems. pp. 283--292. Springer Nature Switzerland, Cham (2024)

\bibitem{NIPS2016_c7635bfd}
Foerster, J., Assael, I.A., de~Freitas, N., Whiteson, S.: Learning to communicate with deep multi-agent reinforcement learning. In: Lee, D., Sugiyama, M., Luxburg, U., Guyon, I., Garnett, R. (eds.) Advances in Neural Information Processing Systems. vol.~29. Curran Associates, Inc. (2016)

\bibitem{9361651}
Ghandar, A., Ahmed, A., Zulfiqar, S., Hua, Z., Hanai, M., Theodoropoulos, G.: A decision support system for urban agriculture using digital twin: A case study with aquaponics. IEEE Access  \textbf{9},  35691--35708 (2021). \doi{10.1109/ACCESS.2021.3061722}

\bibitem{grieves2016digital}
Grieves, M., Vickers, J.: Digital twin: Mitigating unpredictable, undesirable emergent behavior in complex systems. In: Transdisciplinary perspectives on complex systems: New findings and approaches, pp. 85--113. Springer (2016)

\bibitem{10838936}
Hua, Z., Djemame, K., Tziritas, N., Theodoropoulos, G.: A framework for digital twin collaboration. In: 2024 Winter Simulation Conference (WSC). pp. 3046--3057 (2024). \doi{10.1109/WSC63780.2024.10838936}

\bibitem{10.1007/978-3-031-69583-4_18}
Huang, Z., Zhang, N., Shen, J., Diamantopoulos, G., Hua, Z., Tziritas, N., Theodoropoulos, G.: Distributed simulation for digital twins of large-scale real-world diffserv-based networks. In: Carretero, J., Shende, S., Garcia-Blas, J., Brandic, I., Olcoz, K., Schreiber, M. (eds.) Euro-Par 2024: Parallel Processing. pp. 255--269. Springer Nature Switzerland, Cham (2024)

\bibitem{liu2013opportunities}
Liu, C., Chau, K., Wu, D., Gao, S.: Opportunities and challenges of vehicle-to-home, vehicle-to-vehicle, and vehicle-to-grid technologies. Proceedings of the IEEE  \textbf{101}(11),  2409--2427 (2013)

\bibitem{LOTFI2024925}
Lotfi, S., Sedighizadeh, M., Abbasi, R., Hosseinian, S.H.: Vehicle-to-grid bidding for regulation and spinning reserve markets: A robust optimal coordinated charging approach. Energy Reports  \textbf{11},  925--936 (2024). \doi{10.1016/j.egyr.2023.12.044}

\bibitem{lowe2017multi}
Lowe, R., Wu, Y., Tamar, A., Harb, J., Abbeel, P., Mordatch, I.: Multi-agent actor-critic for mixed cooperative-competitive environments. Neural Information Processing Systems (NIPS)  (2017)

\bibitem{ma2012modeling}
Ma, Y., Houghton, T., Cruden, A., Infield, D.: Modeling the benefits of vehicle-to-grid technology to a power system. IEEE Transactions on power systems  \textbf{27}(2),  1012--1020 (2012)

\bibitem{QIU2023120526}
Qiu, D., Xue, J., Zhang, T., Wang, J., Sun, M.: Federated reinforcement learning for smart building joint peer-to-peer energy and carbon allowance trading. Applied Energy  \textbf{333},  120526 (2023). \doi{10.1016/j.apenergy.2022.120526}

\bibitem{sunehag2017value}
Sunehag, P., Lever, G., Gruslys, A., Czarnecki, W.M., Zambaldi, V., Jaderberg, M., Lanctot, M., Sonnerat, N., Leibo, J.Z., Tuyls, K., et~al.: Value-decomposition networks for cooperative multi-agent learning. arXiv preprint arXiv:1706.05296  (2017)

\bibitem{tan2016integration}
Tan, K.M., Ramachandaramurthy, V.K., Yong, J.Y.: Integration of electric vehicles in smart grid: A review on vehicle to grid technologies and optimization techniques. Renewable and Sustainable Energy Reviews  \textbf{53},  720--732 (2016)

\bibitem{10305745}
Vergara, C., Bahsoon, R., Theodoropoulos, G., Yanez, W., Tziritas, N.: Federated digital twin. In: 2023 IEEE/ACM 27th International Symposium on Distributed Simulation and Real Time Applications (DS-RT). pp. 115--116 (2023). \doi{10.1109/DS-RT58998.2023.00027}

\bibitem{43139e3436ca450aae0690bdcd3df024}
{Vergara Marcillo}, C., Bahsoon, R., Tziritas, N., Theodoropoulos, G.: A connectionist approach to federated digital twins. In: Lees, M., Cai, W., Cheong, S., Su, Y., Abramson, D., Dongarra, J., Sloot, P. (eds.) Computational Science – ICCS 2025. pp. 60--74. Lecture Notes in Computer Science, Springer (Jul 2025). \doi{10.1007/978-3-031-97632-2_5}, 25th International Conference on Computational Science (ICCS 2025) ; Conference date: 07-07-2025 Through 09-07-2025

\bibitem{villalonga2020local}
Villalonga, A., Negri, E., Fumagalli, L., Macchi, M., Casta{\~n}o, F., Haber, R.: Local decision making based on distributed digital twin framework. IFAC-PapersOnLine  \textbf{53}(2),  10568--10573 (2020)

\bibitem{7899469}
Yoon, S., Park, K., Hwang, E.: Connected electric vehicles for flexible vehicle-to-grid (v2g) services. In: 2017 International Conference on Information Networking (ICOIN). pp. 411--413 (2017). \doi{10.1109/ICOIN.2017.7899469}

\bibitem{9283357}
Zhang, N., Bahsoon, R., Theodoropoulos, G.: Towards engineering cognitive digital twins with self-awareness. In: 2020 IEEE International Conference on Systems, Man, and Cybernetics (SMC). pp. 3891--3891 (2020). \doi{10.1109/SMC42975.2020.9283357}

\bibitem{10.1145/3635306}
Zhang, N., Bahsoon, R., Tziritas, N., Theodoropoulos, G.: Knowledge equivalence in digital twins of intelligent systems. ACM Trans. Model. Comput. Simul.  \textbf{34}(1) (2024). \doi{10.1145/3635306}

\bibitem{9664795}
Zhang, Y., Sun, J., Wu, C.: Vehicle-to-grid coordination via mean field game. IEEE Control Systems Letters  \textbf{6},  2084--2089 (2022). \doi{10.1109/LCSYS.2021.3139266}

\bibitem{10500853}
Zhou, Z., Li, X., Ge, H., Zhang, J., Xue, Y.: Congestion-aware rebalancing and vehicle-to-grid coordination of shared electric vehicles: An aggregative game approach. IEEE Transactions on Transportation Electrification  \textbf{11}(1),  275--285 (2025). \doi{10.1109/TTE.2024.3389712}

\end{thebibliography}

\end{document}